\documentclass{JAC2003}

\usepackage{graphicx}

\usepackage{amssymb,epsfig,psfrag}

\begin{document}

\onecolumn
\setcounter{page}{0}
{\Large
\strut\hfill PITHA 03/07 \\
\strut\hfill DESY 03-104 \\
\strut\hfill hep-ph/0308094 \\[3cm]
\begin{center}
{\Large{\bf{Top quark physics and QCD: \\[.5em]
Progress since the TESLA TDR}}}%
\footnote{To appear in the proceedings of {\it The 4th ECFA/DESY
Workshop on Physics and Detectors for a 90-800 GeV Linear $e^+e^-$ Collider},
NIKHEF, Amsterdam, the Netherlands, April 2003.}
\\[1cm]
A. Brandenburg \\[1cm]
{\it Institut f\"ur Theoretische 
Physik, RWTH Aachen, D-52056 Aachen, Germany\\
and DESY Theory Group, D-22607 Hamburg,
Germany}
\end{center}\par
\vskip 4cm {\bf Abstract:} \par
I review progress on investigations concerning 
top quark physics and QCD at a future linear $e^+e^-$ 
collider that has been achieved since the presentation of the
TESLA technical design report \cite{TDR} 
in spring 2001. I concentrate on studies that
have been presented during the workshop series of the  
Extended Joint ECFA/DESY Study
on Physics and Detectors for a Linear Electron-Positron Collider. 
\par \vfill \noindent August 2003 \\[1em] \null }
\twocolumn
\clearpage

\title{Top quark physics and QCD: Progress since the TESLA 
TDR\thanks{Much of the work reported in this talk was done by members of the 
Top and QCD working group of the Extended ECFA/DESY Study: 
W. Bernreuther (RWTH Aachen), G.A. Blair (London U.), 
E. Boos (Moscow State U.), 
P.N. Burrows (London U.),
M.P. Casado (CERN), S.V. Chekanov (Argonne), 
S. Dittmaier (MPI M\"unchen), M. Dubinin (Moscow U.), 
J. Fleischer (Bielefeld U.), A. Gay (IRES Strasbourg), 
T. Hahn (MPI M\"unchen), S. Heinemeyer (M\"unchen U.),
A.H. Hoang (MPI M\"unchen), W. Hollik (MPI M\"unchen), 
K. Kolodziej (Silesia U.),
S. Kraml (CERN), F. Krauss (CERN), 
J.H. K\"uhn (Karlsruhe U.), J. Kwiecinski (Inst. of Nucl. Phys. Krakow), 
A. Lorca (DESY Zeuthen), M. Maniatis (Hamburg U.), 
A.V. Manohar (UC San Diego), 
M. Martinez (Barcelona U.), R. Miquel (LBL Berkeley),  
V.L. Morgunov (DESY), W. Porod (Z\"urich U.), T. Riemann (DESY Zeuthen), 
M. Roth (Karlsruhe U.), 
T. Robens (Heidelberg U.), C. Schappacher (Karlsruhe U.), 
I.W. Stewart (MIT), 
C. Sturm (Karlsruhe U.), 
T. Teubner (CERN), P. Uwer (Karlsruhe U.), 
G. Weiglein (IPPP Durham), 
A. Werthenbach (CERN), M. Winter (IRES Strasbourg), 
P.M. Zerwas (DESY Hamburg). I also would like to thank
B.A. Kniehl, A.A. Penin and M. Steinhauser for discussions.}}

\author{A. Brandenburg,
Institut f\"ur Theoretische 
Physik, RWTH Aachen, D-52056 Aachen, Germany\\
and DESY Theory Group, D-22607 Hamburg, \
Germany\thanks{arnd.brandenburg@desy.de}}

\maketitle

\begin{abstract}
I review progress on investigations concerning 
top quark physics and QCD at a future linear $e^+e^-$ 
collider that has been achieved since the presentation of the
TESLA technical design report \cite{TDR} 
in spring 2001. I concentrate on studies that
have been presented during the workshop series of the  
Extended Joint ECFA/DESY Study
on Physics and Detectors for a Linear Electron-Positron Collider. 
\end{abstract}

\section{Introduction}
Since the presentation of the TESLA technical design report \cite{TDR}
in spring 2001, important progress has been achieved and reported in 
the top quark/QCD working group of the  Extended Joint ECFA/DESY Study
on Physics and Detectors for a Linear Electron-Positron Collider.
The common aim of these studies is to improve 
theoretical predictions and perform more realistic simulations  
in order to obtain an accurate understanding of 
top quark interactions and QCD
phenomena at a linear collider. 
A basic issue is a precision determination of two fundamental parameters
of the Standard Model, namely the top quark mass $m_t$ and the strong
coupling constant $\alpha_s$.
These parameters as well as the top quark width can be extracted
from a scan of the $t\bar{t}$ 
threshold cross section with high accuracy, and I will report
on the progress of the simulation
of such a scan and of the refinements in the theoretical computation
of the threshold cross section.
Before that, I will summarize a recent study on the importance
of a very precise measurement of $m_t$. 
Further topics covered here include new studies on 
top quark production and decay in the continuum
and a summary of QCD-related studies. I concentrate on 
work reported at the ECFA/DESY workshops. A summary of top quark and QCD 
studies presented at the last International Linear Collider Workshop (LCWS02)
is given in \cite{Sumino:2003gt}.    
\section{Why do we want to know $m_t$ very precisely?}
\begin{figure}[htb]
\centering
\includegraphics*[width=82mm]{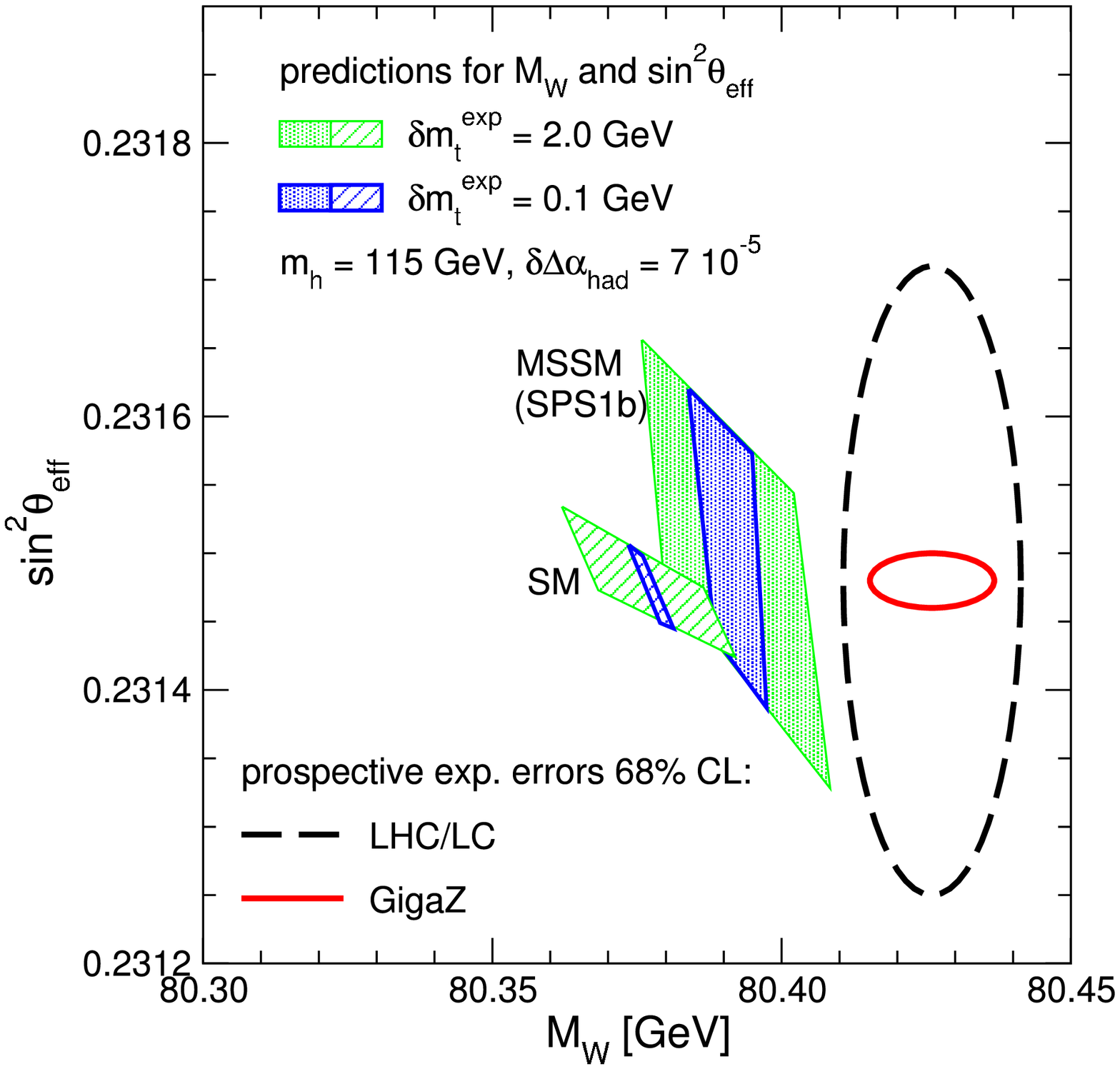}
\caption{The predictions for $M_W$ and $\sin^2\theta_{\rm eff}$
in the SM and the MSSM (SPS1b).
Figure taken from \cite{Hei03}.}
\label{heinemeyer}
\end{figure}

The physics impact of a very precise measurement 
of the top quark mass with $\delta m_t\lesssim 100$ MeV has been
recently studied in detail \cite{Hei03}. 
An accurate knowledge of $m_t$ strongly 
affects tests of the Standard Model (SM) and its extensions
using electroweak precision observables.
This is demonstrated in Fig.~\ref{heinemeyer}, where the prospective
experimental errors of $M_W$ and $\sin^2\theta_{\rm eff}$
at the LHC/LC and the GigaZ option of the LC are compared
to theoretical predictions within the SM and the Minimal Supersymmetric
extension of the SM (MSSM). Since these observables receive 
radiative corrections $\sim m_t^2$, an improvement from $\delta m_t$= 2 GeV
(a value to be obtained at the LHC) to $\delta m_t$= 100 MeV
leads to a significant reduction of the allowed parameter space
both in the SM (about factor of 10) and in the MSSM (about a factor
of 2). This will be very important in the effort to constrain new interactions
in using electroweak precision observables.  
A precise knowledge
of $m_t$ also improves the indirect determination of the top quark Yukawa 
coupling from electroweak precision observables. Further, if 
one wants to obtain constraints on the MSSM
by comparing a precise measurement of the Higgs boson mass with the
theoretical predictions of $m_h$ in this model, 
a precise value of $m_t$ is mandatory due to the strong
dependence ($\sim m_t^4$) of $m_h$ on the top quark mass.
For further details and other applications of a precision 
measurement of $m_t$, see \cite{Hei03}.
\section{Top quark pair production close to threshold}
\subsection{Update of $t\bar{t}$ threshold scan simulation}
\begin{figure}[htb]
\centering
\includegraphics*[width=81mm]{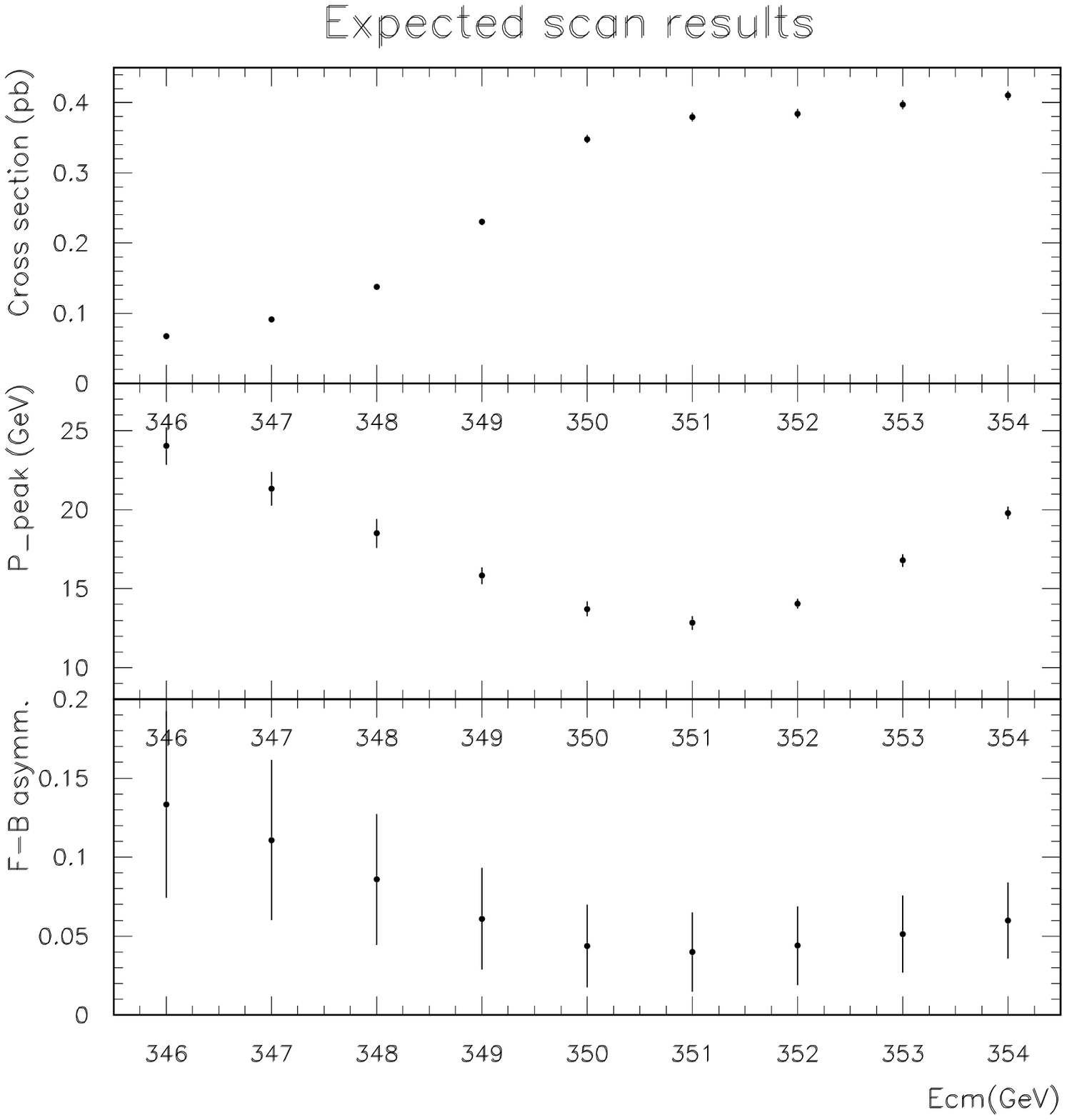}
\caption{Expected scan result for the cross section, the peak of the
top quark momentum distribution and the forward-backward charge asymmetry.
Figure taken from \cite{Mar02}.}
\label{martinez1}
\end{figure}

\begin{figure}[htb]
\centering
\includegraphics*[width=81mm]{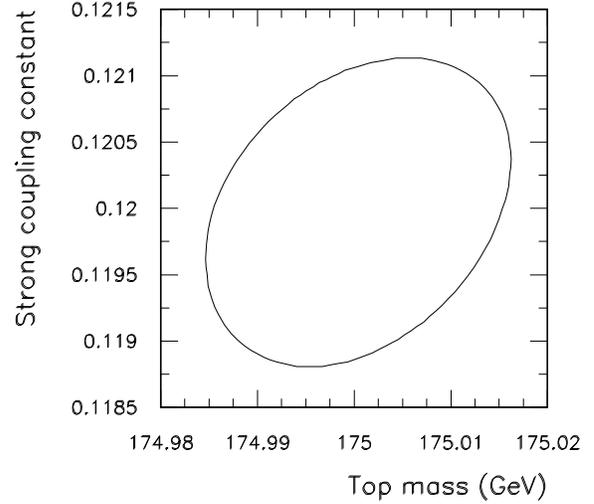}
\caption{$\Delta\chi^2=1$ contour as a function of $m_t^{1S}$ and
$\alpha_s(M_Z)$.
Figure taken from \cite{Mar02}.}
\label{martinez2}
\end{figure}

Recently, an updated $t\bar{t}$ threshold scan simulation has been
performed \cite{Mar02}. It comprises several new features as compared 
to previous studies. First, three observables have been considered: the total
cross section, the position of the peak of the top quark momentum 
distribution, and
the  forward-backward asymmetry. Second, a multiparameter fit with
up to four parameters ($m_t,\ \alpha_s,\ \Gamma_t$ and the top quark Yukawa 
coupling $\lambda_t$) has been performed. Finally, apart from experimental
systematic errors an estimate of the theoretical error in the cross section
prediction has been included in the fits. An integrated luminosity
of ${\cal L}=300{\ \rm pb}^{-1}$ was distributed equally among 10
scan points, where one of them was placed well below threshold in order
to determine directly the background. A theoretical error on the total cross
section of $\Delta\sigma/\sigma=3$\% was assumed in the simulation (see below
for a discussion).
The results may thus give a good impression about the final experimental 
accuracy of the determination of the parameters. The expected scan results
are shown in Fig.~\ref{martinez1}. 
From a two parameter fit for $\alpha_s$ and
$m_t$ the following estimates of their errors were obtained:
\begin{eqnarray}
\Delta m_t^{1S}= 16\ {\rm MeV}, \hspace{2cm} \Delta{\alpha_s}=0.0012.
\end{eqnarray}
Here, $m_t^{1S}$ denotes the $1S$ mass of the top quark, the usage
of which stabilises the location of the threshold with respect to
higher order corrections and reduces the 
correlations between this mass and $\alpha_s$.
The correlation plot between $m_t$ and $\alpha_s$ is shown in 
Fig.~\ref{martinez2}. The correlation coefficient is $\rho=0.33$.
While the cross section has the highest sensitivity on both 
$m_t$ and $\alpha_s$, the additional 
measurement of the  peak of the momentum distribution reduces the
errors and the correlation substantially. 
\par The size of the top quark width $\Gamma_t$ determines how pronounced 
the $1S$ resonance is. A three-parameter fit for $m_t$, $\alpha_s$ and
$\Gamma_t$ gives:
\begin{eqnarray}
\Delta m_t^{1S}= 19\ {\rm MeV}, \hspace{0.1cm} \Delta{\alpha_s}=0.0012,
 \hspace{0.1cm} \Delta{\Gamma_t}=32\ {\rm MeV}.
\end{eqnarray}  
This means that the top quark width can be determined with 2\% accuracy, 
which is a factor of about 9 better than reported in earlier studies.
This improvement is due to assuming a higher integrated luminosity,
a better selection efficiency for $t\bar{t}$ events, a sharper
TESLA beam spectrum and a better scanning strategy when using the $1S$ mass.
\par
The sensitivity of a threshold scan to the top quark Yukawa coupling
$\lambda_t$
through a modification of the $t\bar{t}$ potential 
 is not very large: if one performs a four-parameter fit with an 
external constraint on $\alpha_s(M_Z)$, the results are
(for $M_H=120$ GeV):
\begin{eqnarray}
&& \Delta m_t^{1S}= 31\ {\rm MeV},  \hspace{0.25cm} 
\Delta{\alpha_s}=0.001\ ({\rm constr.}), \nonumber \\ &&
 \Delta{\Gamma_t}=34\ {\rm MeV},  \hspace{0.25cm} 
\frac{\Delta\lambda_t}{\lambda_t}=^{+0.35}_{-0.65}.
\end{eqnarray}
Thus, constraining the top quark Yukawa coupling from a threshold scan is a
 challenging task. A better method is provided by analysing
the associated Higgs production process $e^+e^-\to t\bar{t}H$ \cite{Desch}.
\par
The very accurate measurement of $m_t^{1S}$ is certainly impressive; however,
in order to use the top mass as an input for precision tests of the SM, 
we have to convert the $1S$ mass to  the $\overline{\rm MS}$ mass. The current
theoretical uncertainty in the perturbative relation between these two masses
is of the order of 100 MeV \cite{Hoa00}. 
\begin{figure*}
  \unitlength1.0cm
  \begin{center}
    \begin{picture}(15,15)
      \put(-2.8,2){\psfig{figure=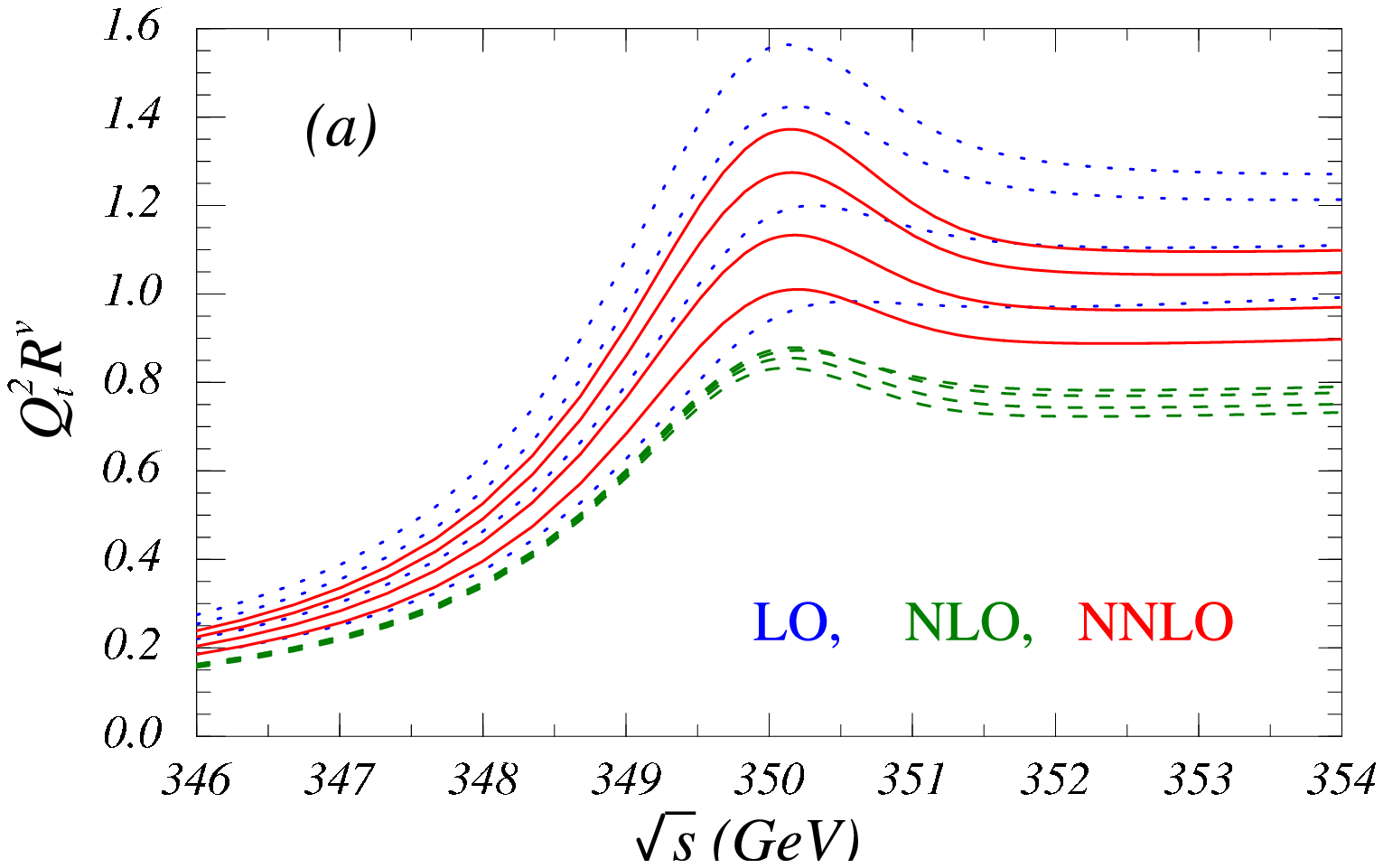,width=10cm}}
      \put(5.6,2){\psfig{figure=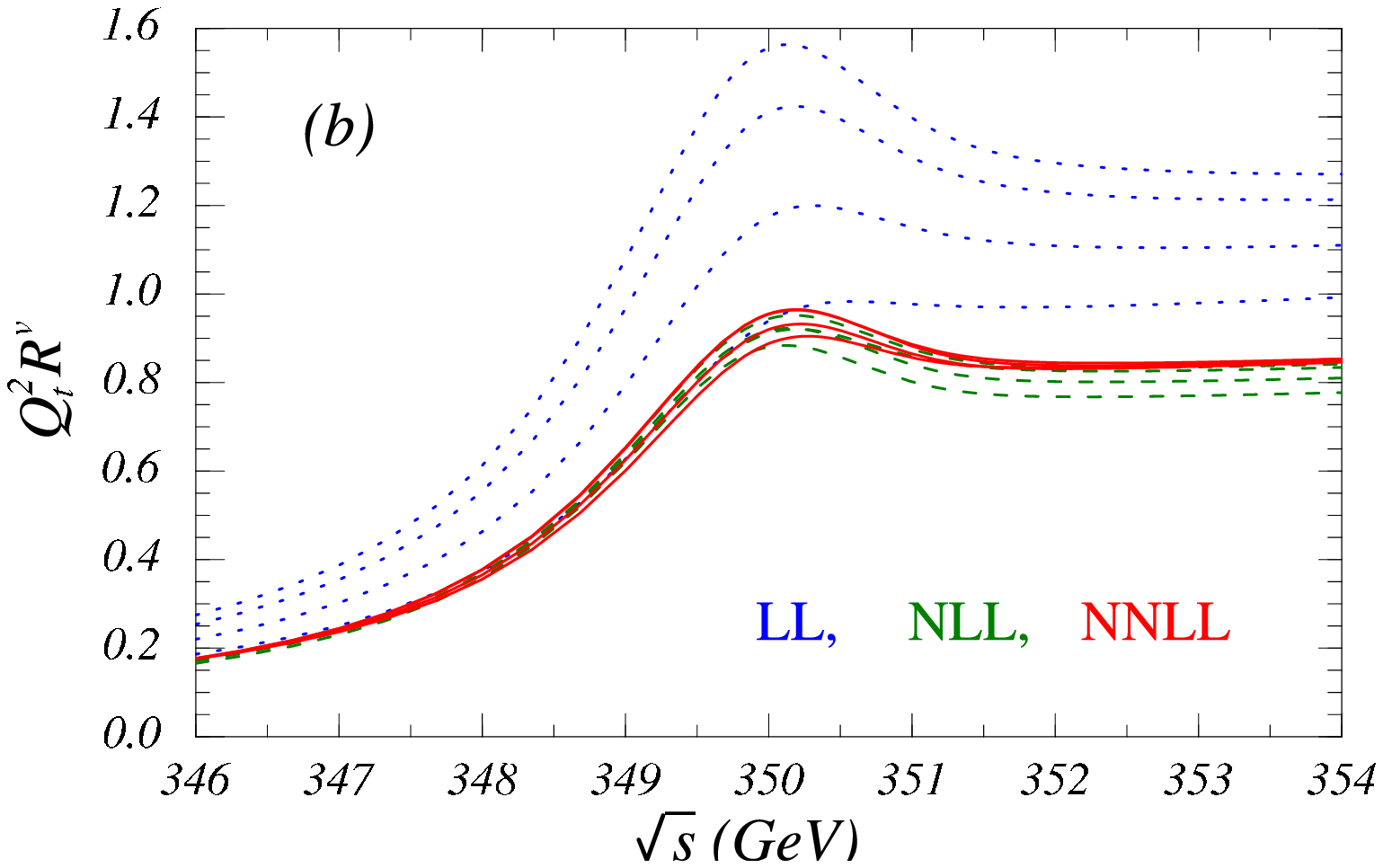,width=10cm}}
    \end{picture}
\\[-10.5cm]
\caption{Results for the vector current $R$-ratio with fixed $m_t^{1S}$
mass for fixed order and renormalization group improved predictions. The dotted, dashed, and solid curves in a) are LO, NLO, and NNLO, and in b) are LL, NLL, and NNLL order.
For each order four curves are plotted for velocity renormalization scales 
$\nu=0.1,\ 0.125,\ 0.2$ and 0.4. 
Figure taken from \cite{Hoa01} (an update with very small changes
was given in \cite{Hoa02}).}
\label{hoang1}
  \end{center}
\end{figure*}
\subsection{Theoretical developments}
The status of $t\bar{t}$ threshold cross section 
calculations in spring 2001 was as 
follows: Several groups had calculated the cross section at NNLO (for a 
synopsis of these results and further references, see \cite{Hoa00}).
The corrections turned out to be large, and the threshold location was
found to be
unstable under perturbative corrections when the top quark pole mass was 
used in the calculation. Further, a strong correlation between
$\alpha_s$ and $m_t$ limited the experimental precision of $m_t$ to about
300 MeV. The usage of threshold masses \cite{Ben98,Hoa99,Big97} 
reduced this correlation and stabilized the
position of the threshold 
significantly. However, the height of the cross section 
still suffered from
large perturbative corrections of the order of 20 to 30 \%, even when expressed
in terms of a top quark threshold mass. In order to improve the prediction
of the threshold cross section, the impact of a 
summation of QCD logarithms 
of ratios of the scales $m_t$, $m_tv$, and $m_tv^2$
was computed in \cite{Hoa01,Pin01}. 
A comparison of the fixed order results and 
the renormalization group improved
results
is shown in Fig.~\ref{hoang1}.
The remaining theoretical uncertainty
of the cross section was estimated in \cite{Hoa01} to be $\pm 3\%$.
This number was obtained by varying the dimensionless
velocity subtraction scale that separates hard, soft and ultrasoft
momenta and by estimating the size of the one yet unknown NNLL
contribution from the running at the production current.
Very recently, the NNLL non-mixing contributions to the running of the
production current have been determined \cite{Hoa03}. It remains to
be seen whether the $\pm 3\%$ estimate will withstand future
refinements of the  cross section 
calculations. For testing the
convergence of the alternative fixed order (LO, NLO, NNLO, \ldots) 
perturbation series the
computation of the NNNLO contributions is mandatory.  Important
progress has been recently achieved in this direction \cite{Kni02}.
\par
Electoweak effects have not yet been consistently included either
at NNLL order or NNLO.
\section{Top quark production and decay in the continuum}
\subsection{Mass determination from continuum production}
\begin{figure}[htb]
\centering
\includegraphics*[width=82mm]{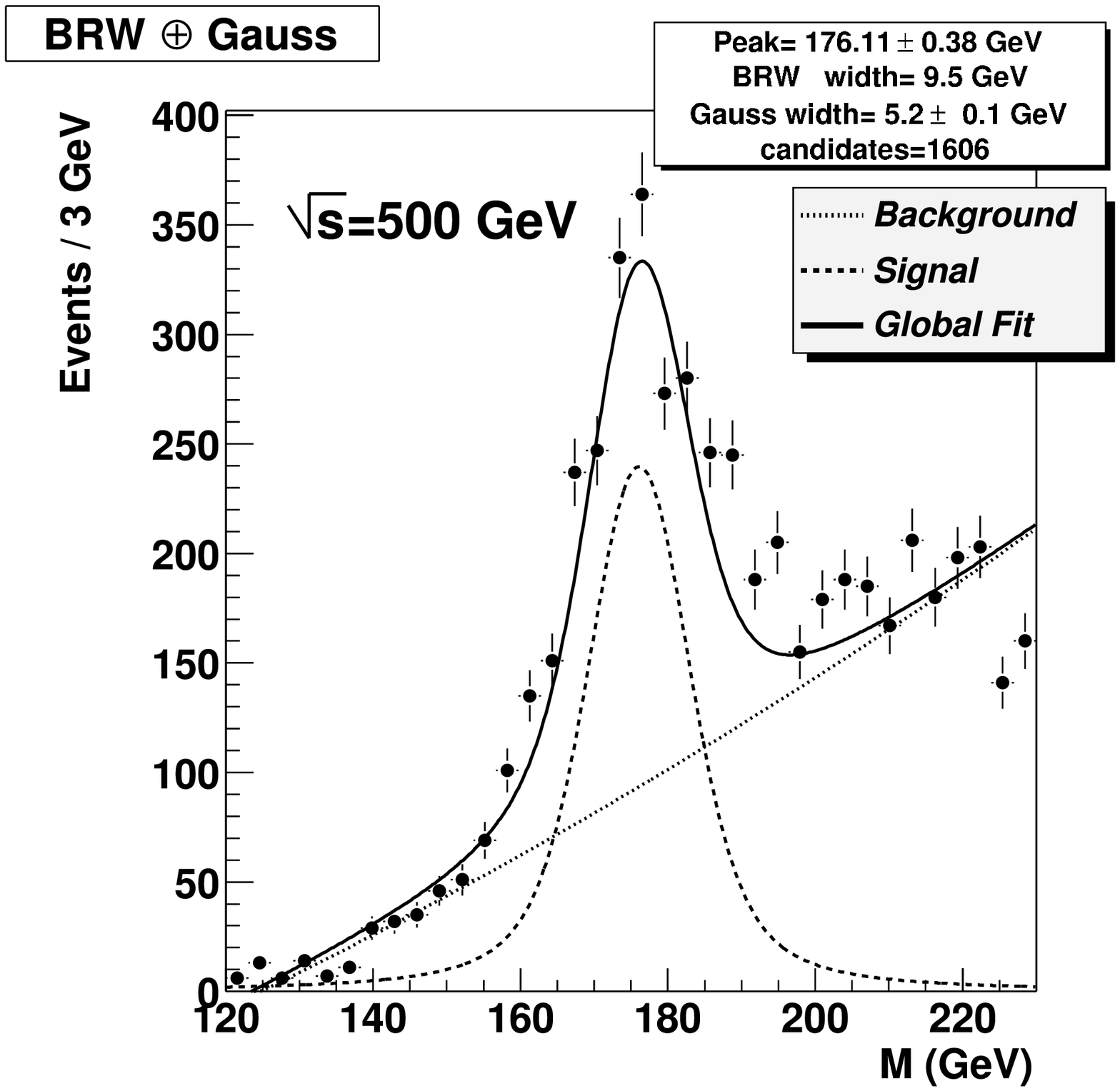}
\caption{The invariant-mass distribution for three-jet clusters
in $e^+e^-\to 6$ jets.
Figure taken from \cite{Che03}.}
\label{morgunov}
\end{figure}

In \cite{Che03} the possibility of measuring the top quark mass 
in the continuum was investigated. The process $e^+e^-\to t\bar{t}\to 6$ jets
was simulated including the QCD background. Top quarks were reconstructed
by grouping the 6 jets into pairs of three-jet groups. 
Only three-jet groups which are
produced back-to-back are accepted. The three-jet invariant mass distribution
(see Fig.~\ref{morgunov}) then shows a prominent peak, the position of  
which is interpreted as the top quark mass.
The statistical uncertainty of the peak position is 100 MeV for an integrated
luminosity of 300 ${\rm pb}^{-1}$ at $\sqrt{s}=500$ GeV. Experimental
systematic errors have not yet been studied. Further, it is not clear yet
how to relate this `kinematic mass` to the pole mass or other top quark 
mass definitions. The method was recently extended to 
semileptonic top decays \cite{Mor03}. 
\subsection{Anomalous top quark couplings}
A new analysis was started to evaluate the 
sensitivity of $e^+e^-\to t\bar{t}$ to anomalous top quark couplings
\cite{Cas02}. The plan is to use PANDORA/PYTHIA  and SIMDET
in the simulation and try to find 
observables that optimize the sensitivity.
\subsection{New theoretical studies and tools}
In the following I briefly discuss further studies on top quark
production and decay in the continuum that
have been presented during the workshop series.
\begin{itemize}
\item[$\bullet$] Several  calculations of the electroweak one-loop 
radiative corrections to the process 
$e^+e^-\to t\bar{t}$ have been compared in detail \cite{Fle02a}.
The numerical agreement is excellent (see also Table 4 in
\cite{Dit03}). The package {\tt topfit} 
containing these
corrections is publicly available \cite{topfit}. 
\item[$\bullet$] New tree level Monte Carlo generators 
({\tt AMEGIC++} \cite{Kra02}, {\tt eett6f} 
\cite{Kol01} and {\tt LUSIFER} \cite{Dit02}) for the processes 
$e^+e^-\to 6$ fermions have been written (for details, see
\cite{Dit03}).
These programs  allow to study in particular the non-resonant background
to $t\bar{t}$ production and decay.    
\item[$\bullet$] In \cite{Boo01}, the production of single top quarks
in $e^+e^-,e^-e^-,e\gamma$ and $\gamma\gamma$ collisions was studied
at tree level. 
By comparing all possible reactions, the best option turned out to be 
collisions of circular polarized photons with left-handed electrons.
The cross section at $\sqrt{s}=500$ GeV is $\sigma(\gamma_+e^-_L\to \bar{t}
b\nu)\sim 100$ fb, and this process is very sensitive to $V_{tb}$ as well
as to possible anomalous couplings. If one aims at an experimental
precision of 1\% for $V_{tb}$, the inclusion of higher order
corrections is mandatory. The QCD  
corrections to this process have been computed very recently \cite{Kuh03}
and are of the order of 5\%.
\item[$\bullet$] The SUSY-QCD corrections
to the production and decay of {\it polarized} top quarks 
in $e^+e^-$ collisions have been computed in \cite{Bra02}. While
the decay width and lepton energy spectrum can be modified at the
percent level, top polarization observables are hardly affected by these
corrections.
\end{itemize}
\section{QCD studies}
\subsection{Measurement of $\alpha_s$}
The primary goal of QCD studies at a linear collider is 
to measure the strong
coupling constant $\alpha_s$ as precisely as possible.
The aim is to reduce the current accuracy $\Delta\alpha_s(M_Z)=0.003$ to 
a value of $\Delta\alpha_s(M_Z)=0.001$ or smaller.
In the context of QCD, such an accuracy is important, since {\it all}
predictions of perturbative QCD are directly affected, in particular
multi-jet cross sections at higher orders. Furthermore, an extrapolation
of $\alpha_s(Q)$ to very high energy scales which is performed to test
the hypothesis of Grand Unification needs precise initial conditions, and
the uncertainty on $\alpha_s$ is currently the limiting factor of such  
tests. This is illustrated in Fig.~6, where the running of the inverse
coupling constants is shown. The narrow error band on $1/\alpha_3$ in
Fig.~6b corresponds to  $\Delta\alpha_s(M_Z)=0.001$.
The techniques for a determination of $\alpha_s(M_Z)$ at TESLA have been
described in detail in the TDR. In a recent study  \cite{Win01}, the prospects 
of a measurement of  $\alpha_s$ from GigaZ
analyses have been investigated. The factor of $\sim 100$ 
in the size of the data
sample as compared to LEP data together with the expected better performance
of the detector give rise to the expectation that systematic errors
may shrink by a factor of 3 to 5. This would mean that the experimental 
accuracy on $\alpha_s$ could be brought down to $(5-7)\times 10^{-4}$, 
the most sensitive observable being the inclusive ratio 
$\Gamma_Z^{\rm hadron}/\Gamma_Z^{\rm lepton}$. No theoretical errors
are included in this analysis. 
At present, the theoretical uncertainty of $\alpha_s$-determinations from 
$\Gamma_Z^{\rm hadron}/\Gamma_Z^{\rm lepton}$ is estimated to be
of the same size as the current experimental accuracy \cite{Bet00}. 
In view of the prospective accuracy from a GigaZ run, there is an 
ongoing effort to compute more and more terms of the perturbation series for
 $\Gamma_Z^{\rm hadron}/\Gamma_Z^{\rm lepton}$
and related quantities like $R(s)$ 
and $\Gamma_{\tau}^{\rm hadron}/\Gamma_{\tau}^{\rm lepton}$. 
The most recent step in this direction 
has been the calculation of a gauge-invariant subset 
of the order $\alpha_s^4$ contributions,
namely the terms of order $\alpha_s^4n_f^2$, where $n_f$ is the number of
fermion flavours \cite{Bai02}.   

The bottleneck of determinations of $\alpha_s(M_Z)$ from event shapes
(like thrust distribution, jet rates etc.)
is currently the insufficient theoretical precision of perturbative
QCD calculations. 
Currently most shape variables 
are known to next-to-leading order
accuracy, while for some observables resummed calculations are available.
However, enormous progress towards the calculation of 
$e^+e^-\to 3$ jets to NNLO (${\cal O}(\alpha_s^3)$) has been achieved
within the last few years \cite{nnlo}. It is estimated that once these
calculations are accomplished, the current theoretical uncertainty  
(obtained by a variation of the QCD renormalisation scale) of 
$\Delta\alpha_s(M_Z)\simeq 0.006$
will shrink by a factor of 3 to 5.   
\begin{figure*}[t!]
\setlength{\unitlength}{1mm}
\begin{center}
\begin{picture}(160,85)
\put(-14,-97){\mbox{\epsfig{figure=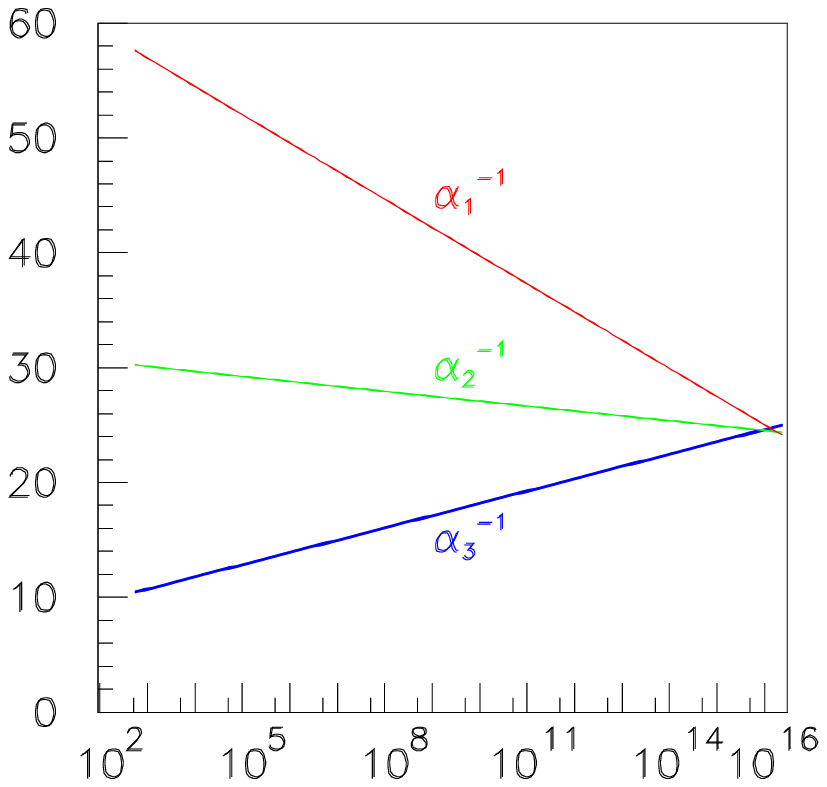,
                                   height=19.cm,width=19.5cm}}}
\put(84,-3){\mbox{\epsfig{figure=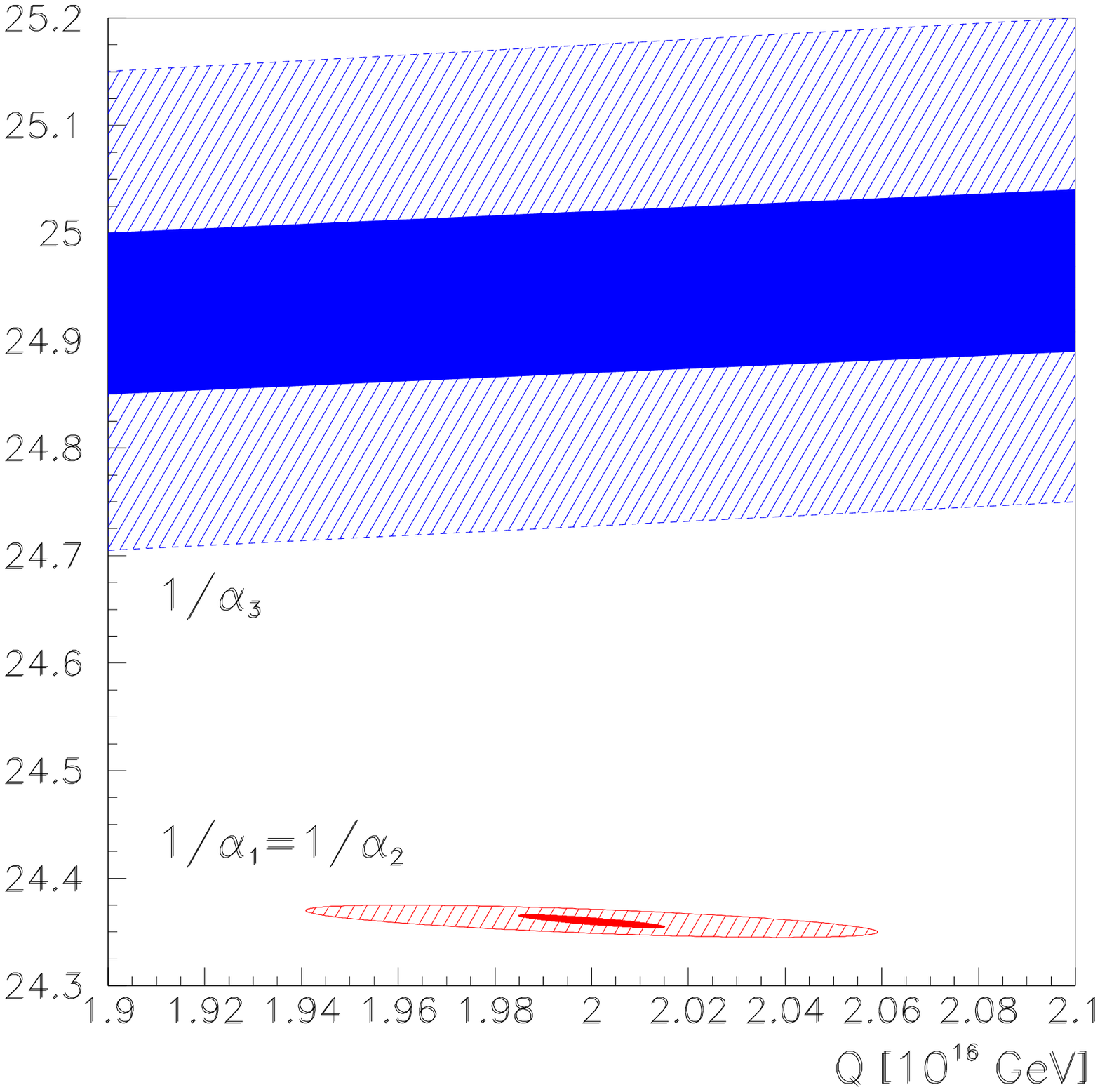,
                                   height=8.5cm,width=8.5cm}}}
\put(-3,80){\mbox{\bf a)}}
\put(84,80){\mbox{\bf b)}}
\put(60,-3){\mbox{Q~[GeV]}}
\end{picture}
\end{center}
\caption{\it a) Running of the inverse gauge couplings. b) Determination 
             of $M_U$, $\alpha_U^{-1}$; the unification point $U$ is
             defined by the meeting point of $\alpha_1$ with $\alpha_2$.
             The wide error bands are based on present data, the narrow
             bands demonstrate the improvement expected by future GigaZ 
             analyses. Figure taken from \cite{Bla02}.}
\label{fig:gauge}
\end{figure*} 

\subsection{Saturation model for $\gamma\gamma$ and $\gamma^*\gamma^*$ 
processes}
In \cite{Tim01} a saturation model has been constructed 
to describe the total cross
section for $\gamma\gamma$ and $\gamma^*\gamma^*$ collisions at high energies.
The $\gamma^*\gamma^*$ total  cross section is assumed to be dominated
by interactions of two colour dipoles into which the photons can fluctuate,
the novel feature being that the saturation property of the dipole-dipole cross
section is incorporated. This allows to describe the variation of the
energy dependence 
of the cross section when the virtualities of the photons change. The model
fits the available two-photon data reasonably well, except for 
$b$-quark production. Predicitions for TESLA energies have been 
formulated. 
\subsection{Odderon contribution to exclusive pion-photoproduction}
In \cite{Rob03} the nonperturbative odderon contributions to the
processes $e^+e^-,\gamma\gamma\to \pi^0\pi^0$ 
have been studied. The cross sections are very sensitive, 
both in the $e^+e^-$ and the
$\gamma\gamma$ mode, to the coupling 
strength of the odderon, while the Regge trajectory parameters are harder
to determine. 
\section{Conclusions}
Without doubt a future high-energy, high luminosity $e^+e^-$ linear collider
like TESLA will be an excellent facility to study in detail the
physics of top quarks and strong interaction phenomena.
In particular, a $t\bar{t}$ threshold scan will provide
a measurement of $m_t$ with unmatched precision, with important
implications for searches for physics beyond the Standard Model. 
Important progress
has been achieved recently to predict the threshold cross section
precisely and to evaluate the experimental sensitivity to $m_t$ and
other SM parameters. However, future improvements 
are necessary and possible: 
For example, the experimental uncertainty with which a 
threshold mass like $m_t^{1S}$ can be extracted 
is expected to be much smaller than the current theoretical uncertainty
that relates this parameter to the $\overline{\rm MS}$ mass.
An ongoing challenging task is the refinement of the theoretical
understanding of the threshold cross section.
Furthermore, it  would be desirable to include a realistic luminosity spectrum
$d{\cal L}/dE$ in the threshold scan simulation.
Another important task is a simulation of a determination
of top quark form factors at the detector level.

Challenging issues in perturbative QCD
are the calculation of event shape variables at NNLO
and a further improvement of the theoretical predictions
of inclusive quantities like $\Gamma_Z^{\rm hadron}/\Gamma_Z^{\rm lepton}$. 
Progress in these areas would 
contribute significantly to the aim of determining
$\alpha_s(M_Z)$ with an accuracy of a percent or better.



\begin{thebibliography}{99} 
%
\bibitem{TDR}
J.A.~Aguilar-Saavedra et al., ``TESLA technical design report part III: 
Physics at an $e^+e^-$ linear collider'',  
 hep-ph/0106315, see http://tesla.desy.de/tdr/ .
%
\bibitem{Sumino:2003gt}
Y. Sumino,
Talk presented at LCWS02, Jeju Island, Korea, August 2002, hep-ph/0302086.
%
\bibitem{Hei03}
S. Heinemeyer, S. Kraml, W. Porod and G. Weiglein,
LC-TH-2003-052, hep-ph/0306181.
%
\bibitem{Mar02}
M. Martinez and R. Miquel,
Eur. Phys. Jour. {\bf C 27} (2003) 49 [hep-ph/0207315].
%
\bibitem{Desch} K. Desch, these proceedings.
%
\bibitem{Hoa00} A.H. Hoang et al., Eur. Phys. J. direct C {\bf 3}
(2000) 1 [hep-ph/0001286].
%
\bibitem{Ben98} M. Beneke, Phys. Lett. B {\bf 434} (1998) 115 [hep-ph/9804241].
%
\bibitem{Hoa99} A.H. Hoang and T. Teubner, Phys. Rev. D {\bf 60} 
(1999) 114027 [hep-ph/9904468].
%
\bibitem{Big97} I.I. Bigi, M. Shifman, N. Uraltsev and A. Vainshtein,
Phys. Rev. D {\bf 56} (1997) 4017 [hep-ph/9704245].
%
%
\bibitem{Hoa01} A.H. Hoang, A.V. Manohar, I.W. Stewart and T. Teubner,
Phys. Rev. Lett. {\bf 86} (2001) 1951 [hep-ph/0011254] and
Phys. Rev. D {\bf 65} (2002) 014014 [hep-ph/0107144].
%
\bibitem{Pin01} A. Pineda, Phys. Rev. D {\bf 65} 
(2002) 074007 [hep-ph/0109117] and
Phys. Rev. D {\bf 66} (2002) 054022 [hep-ph/0110216].
%
\bibitem{Hoa02} A.H. Hoang and I.W. Stewart,
Phys. Rev. D {\bf 67} (2003) 114020 [hep-ph/0209340].
%
\bibitem{Hoa03} A.H. Hoang, hep-ph/0307376.
%
\bibitem{Kni02}
B.A. Kniehl, A.A. Penin, V.A. Smirnov and M. Steinhauser,
Nucl. Phys. B  {\bf 635} (2002) 357 [hep-ph/0203166];\\
A.A. Penin and M. Steinhauser, Phys. Lett. B {\bf 538} 
(2002) 335 [hep-ph/0204290];\\
B.A. Kniehl, A.A. Penin, V.A. Smirnov and M. Steinhauser,
Phys. Rev. Lett. {\bf 90} (2003) 212001  [hep-ph/0210161];\\
Y. Kiyo and Y. Sumino, 
Phys.Rev. D {\bf 67} (2003) 071501 [hep-ph/0211299].
%
\bibitem{Che03}
S.V. Chekanov, V.L. Morgunov,
LC-PHSM-2003-001 [hep-ex/0301014].
%
\bibitem{Mor03} S.V. Chekanov, V.L. Morgunov, 
``Top quark at a linear collider'', 
talk presented in the Top and QCD working group session 
at the Fourth ECFA/DESY Workshop
of the Extended Joint ECFA/DESY Study on 
Physics and Detectors for a Linear Electron-Positron Collider
, Amsterdam, The Netherlands, 1.-4. April 2003.
%
\bibitem{Cas02} M.P. Casado, talks  presented in the Top and 
QCD working group session at the First and Second Workshop
of the Extended Joint ECFA/DESY Study on 
Physics and Detectors for a Linear Electron-Positron Collider,
Krakow, Poland, 14.-18. September 2001 and St. Malo, France, 12.-15. April
2002. 
%
\bibitem{Fle02a}
J. Fleischer, T. Hahn, W. Hollik, T. Riemann, C. Schappacher and 
A. Werthenbach,
LC-TH-2002-002, hep-ph/0202109;\\
J. Fleischer, 
J. Fujimoto, T. Ishikawa, A. Leike, T. Riemann, Y. Shimizu
and A. Werthenbach,
KEK Proceedings 2002-11
(2002) (Y. Kurihara, ed.) 153 [hep-ph/0203220];\\
%
T. Hahn, W. Hollik, A. Lorca, T. Riemann and
A. Werthenbach,
hep-ph/0307132.
%
\bibitem{Dit03} S. Dittmaier, these proceedings.
%
\bibitem{topfit} See http://www-zeuthen.desy.de/$\sim$riemann/.
%
\bibitem{Kra02} F. Krauss, R. Kuhn and G. Soff, JHEP {\bf 0202} (2002)
044 [hep-ph/0109036];\\
A. Schalicke, F. Krauss, R. Kuhn and G. Soff, JHEP {\bf 0212} (2002) 013 
[hep-ph/0203259].
%
\bibitem{Kol01} K. Kolodziej, Eur. Phys. J. C {\bf 23} (2002) 471 
[hep-ph/0110063].
\bibitem{Dit02} S. Dittmaier and M. Roth, Nucl. Phys. B {\bf 642}
(2002) 307 [hep-ph/0206070].
%
\bibitem{Boo01} E. Boos, M. Dubinin, A. Pukhov, M. Sachwitz and 
 H.J. Schreiber,
Eur. Phys. J. C {\bf 21} (2001) 81 [hep-ph/0104279].
%
\bibitem{Kuh03} J.H. K\"uhn, C. Sturm, P. Uwer,
hep-ph/0303233.
\bibitem{Bra02} A. Brandenburg and M. Maniatis, 
Phys. Lett. B {\bf 545} (2002) 139 [hep-ph/0207154] and
Phys. Lett. B {\bf 558} (2003) 79 [hep-ph/0301142].
\bibitem{Bla02} G.A. Blair, W. Porod, P.M. Zerwas, Eur. Phys. J. C {\bf 27}
(2003) 263, LC-TH-2003-021 [hep-ph/0210058].
\bibitem{Win01} M. Winter, ``Determination of the strong coupling
constant at GigaZ'', LC-PHSM-2001-016.
%
\bibitem{Bet00} S. Bethke, J. Phys. G {\bf 26} (2000) R27 [hep-ex/0004021].
\bibitem{Bai02} P.A. Baikov, K.G. Chetyrkin, J.H. K\"uhn, 
Phys. Rev. Lett. {\bf 88} (2002) 012001 [hep-ph/0108197].
%
\bibitem{nnlo} L.W. Garland, T. Gehrmann, E.W.N. Glover, A. Koukoutsakis
and E. Remiddi, Nucl. Phys. B {\bf 626} (2002) 107 [hep-ph/0112081] and
Nucl. Phys. B {\bf 642} (2002) 227 [hep-ph/0206067];\\
S. Moch, P. Uwer, S. Weinzierl, Phys. Rev. D {\bf 66} (2002) 114001
[hep-ph/0207043];\\
S. Weinzierl, JHEP 0303 (2003) 062 [hep-ph/0302180] and hep-ph/0306248.
\bibitem{Tim01} N. Timneanu, J. Kwiecinski, L. Motyka,
 Eur. Phys. J. C {\bf 23} (2002) 513 [hep-ph/0110409].
\bibitem{Rob03} T. Robens,
hep-ph/0302048 and talk presented in the Top and QCD working group session 
at the Fourth ECFA/DESY Workshop
of the Extended Joint ECFA/DESY Study on 
Physics and Detectors for a Linear Electron-Positron Collider
, Amsterdam, The Netherlands, 1.-4. April 2003.
\end{thebibliography}
\end{document}